\documentclass[fleqn,usenatbib]{mnras}

\usepackage{newtxtext,newtxmath}

\usepackage[T1]{fontenc}
\DeclareRobustCommand{\VAN}[3]{#2}
\let\VANthebibliography\thebibliography
\def\thebibliography{\DeclareRobustCommand{\VAN}[3]{##3}\VANthebibliography}

\usepackage{graphicx}	
\usepackage{amsmath}	

\defcitealias{Mathis77}{MRN}
\defcitealias{huang2021evolution}{H21}
\defcitealias{hirashita2020spectral}{HDM20}

\title[Dust SEDs in a cosmological simulation]{Dust SEDs in Milky Way-like galaxies in the IllustrisTNG simulations based on the evolution of grain size distribution}

\author[C.-Y. Chang et al.]{
Chiung-Yin Chang,$^{1,2,3}$\thanks{E-mail: cychang@asiaa.sinica.edu.tw}
Yu-Hsiu Huang,$^{2,4}$
Hiroyuki Hirashita$^{2,5}$\thanks{E-mail: hirashita@asiaa.sinica.edu.tw} and
Andrew P. Cooper$^{3,5,6}$
\\
$^{1}$Department of Engineering and System Science, National Tsing Hua University, 101, Section 2, Kuang-Fu Road, Hsinchu 30014, Taiwan \\
$^{2}$Institute of Astronomy and Astrophysics, Academia Sinica,
Astronomy-Mathematics Building, No.\ 1, Section 4,
Roosevelt Road, Taipei 10617, Taiwan\\
$^{3}$Institute of Astronomy and Department of Physics, National Tsing Hua University, 101, Section 2, Kuang-Fu Road, Hsinchu 30014, Taiwan\\
$^{4}$Institute of Physics, National Taiwan University, No.\ 1, Section 4, Roosevelt Road, Taipei 10617, Taiwan\\
$^{5}$Physics Division, National Center for Theoretical Sciences, Taipei 10617, Taiwan\\
$^{6}$Center for Informatics and Computation in Astronomy, National Tsing Hua University, 101, Section 2, Kuang-Fu Road, Hsinchu 30013, Taiwan
}

\date{Accepted XXX. Received YYY; in original form ZZZ}

\pubyear{2022}

\begin{document}
\label{firstpage}
\pagerange{\pageref{firstpage}--\pageref{lastpage}}
\maketitle

\begin{abstract}

To understand how the evolution of grain size distribution in galaxies affects observed dust properties, we apply a post-processing dust evolution model to galaxy merger trees from the IllustrisTNG cosmological hydrodynamical simulation. 
Our dust model includes stellar dust production, sputtering in hot gas, dust growth by accretion and coagulation in the dense interstellar medium (ISM), and shattering in the diffuse ISM. We decompose the grain size distribution into different dust species depending on the elemental abundances and the dense ISM fraction given by the simulation.
In our previous work, we focused on Milky Way (MW) analogs and reproduced the observed MW extinction curve.
In this study, we compute dust spectral energy distributions (SEDs) for the MW analogues.
Our simulated SEDs broadly reproduce the observed MW SED within their dispersion and so does the observational data of nearby galaxies, although they tend to underpredict the MW SED at short wavelengths where emission is dominated by polycyclic aromatic hydrocarbons (PAHs). We find that metallicity and dense gas fraction are the most critical factors for the SED shape, through their influence on coagulation and shattering.
The overall success of our models in reproducing the MW SED further justifies the dust evolution processes included in the model and predicts the dispersion in the SEDs caused by the variety in the assembly history.
We also show that the most significant increase in the dust SED occurs between redshifts $z\sim 3$ and 2 in the progenitors of the simulated MW-like galaxies.
\end{abstract}

\begin{keywords}
methods: numerical -- dust, extinction -- Galaxy: evolution -- galaxies: evolution -- galaxies: ISM
-- infrared: galaxies.
\end{keywords}


\section{Introduction}

Dust has several important effects on galaxy evolution.
Dust absorbs ultraviolet (UV) and optical radiation from stars and re-emits the absorbed energy in the infrared (IR; e.g.\ \citealt{Desert90,Takeuchi05}), greatly altering the spectral energy distributions (SEDs) of galaxies. Fits to observed SEDs are used to extract fundamental properties such as stellar mass, star formation rate, dust mass, and age \citep[e.g.][]{daCunha08,Boquien19}. In such fits it is crucial to account for modifications of the SED by dust extinction and reemission.
IR dust emission is also a good indicator of the star formation rate (SFR) \citep[e.g.][]{Buat96,Kennicutt:1998a,Inoue00}.
Dust also has an important effect on the chemical properties of the interstellar medium (ISM). It is an efficient catalyst for the formation of molecular hydrogen \citep{Gould63,Cazaux04}, the most abundance molecule in star-forming clouds.
Dust also causes depletion of heavy elements from the gas phase \citep[e.g.][]{Savage96,Jenkins09}. This reduces the physical effects of metals, such as metal line absorption and radiative cooling.
Because of these important effects, it is crucial to understand the origin and evolution of dust in the broader context of galaxy formation.

Cosmological hydrodynamic simulations, which solve the gravitational evolution of dark matter together with baryonic processes \citep[e.g.][]{Springel03,Pillepich18a}, have been used to explore how the abundance and properties of dust evolve alongside other galaxy properties.
Direct cosmological simulations are useful because the evolution of dust is strongly coupled to that of the ISM as a whole \citep[e.g.][]{Dwek98}. Previous studies of dust in cosmological simulations include \citet{McKinnon16,McKinnon17}, \citet{Aoyama18} and \citet{Hou19}.  Some have focused on single isolated galaxies \citep[e.g.][]{Bekki13,Aoyama17,Hou17,Osman20},
or on specific regions of the ISM \citep[e.g.][]{Zhukovska16,Hu19}.
Some have performed simulations to reveal the dust properties in high-redshift galaxies \citep[e.g.][]{Graziani20} or in a wide range of redshifts \citep[e.g.][]{Li19}.
Alongside hydrodynamic simulations, semi-analytic models are also useful means of exploring dust evolution in a cosmological context \citep[e.g.][]{Popping17,Vijayan19,Triani20}.

The above-mentioned effects of dust on galaxy evolution depend not only on the total dust amount but also on the total dust surface area \citep[e.g.][]{Yamasawa11,Chen18}
or on the cross-section of grains for interaction with photons. 
These two quantities are determined by the distribution function of dust grain sizes (we refer to this as the grain size distribution). Therefore, it is important to
understand the evolution of the grain size distribution as well as the dust abundance.
There have been some attempts to compute the evolution of the grain size distribution in
cosmological simulations. Since calculating the grain size distribution is computationally
expensive, one approach has been to represent the continuous range of grain sizes with two broad bins
\citep{Aoyama18,Gjergo18,Hou19,Granato21}.
\citet{McKinnon18} developed a detailed framework to calculate the evolution of the full grain size distribution in hydrodynamic simulations. Through tests using simulations of isolated galaxies, they confirmed that the grain size distribution and the total dust mass are strongly affected by multiple dust processing mechanisms, especially shattering and coagulation.
\citet{Aoyama20} included a model for grain size distribution in a simulation of an isolated galaxy
and showed that the grain size distribution is strongly affected by the evolving conditions
(especially the density)
of the ISM. Their work demonstrated that it is important to calculate
the grain size distribution in a
manner consistent with the prevailing physical condition of the ISM.
The above simulations are also used to predict observational dust properties such as extinction curves, as we discuss below.


The grain size distribution is imprinted on a number of observable properties of galaxies.
Among these, extinction curves, which describe the wavelength dependence of
dust extinction, are often used to extract the grain size distribution
\citep[][hereafter \citetalias{Mathis77}]{Mathis77}. In particular, fits to extinction curves in the MW and nearby galaxies have been used to constrain grain size distributions
\citep[e.g.][]{Pei92,Kim94,Weingartner01}. However, whether or not these grain size distributions
are consistent with theories of dust evolution in the
ISM is still to be investigated. \citet{Odonnell97} considered grain coagulation and shattering
in the turbulent ISM to predict the grain size distribution in the MW.
They succeeded in obtaining extinction curves similar to that of the MW.
\citet{Asano14},
based on their model of processes affecting the grain size distribution \citep{Asano13},
predicted the evolution of extinction curve in a galaxy over its full history.
\citet{Nozawa15} extended this model to include dust processing in dense molecular gas and reproduced
the MW extinction curve.

A similar approach has been applied to hydrodynamic simulations to predicted the evolution of the extinction curve, in particular for MW-like galaxies. \citet{Aoyama20}, mentioned above, used a simulation of an isolated disc galaxy similar to the MW. They broadly reproduced the MW extinction curve, finding significant variation within the galaxy.
In effect using the framework developed by \citet{McKinnon18},
\citet{Li21} performed a cosmological simulation and showed that galaxies with mass comparable to the MW have MW-like extinction curves.

Since the direct calculation of grain size distributions within a hydrodynamic simulation is computationally expensive,
\citet[hereafter \citetalias{huang2021evolution}]{huang2021evolution} developed a more efficient post-process approach, which they applied to The Next Generation Illustris project (hereafter IllustrisTNG or TNG, described in Section \ref{sec:sample}). Their dust evolution model was developed separately by
\citet{Hirashita20}.
\citetalias{huang2021evolution} succeeded in reproducing the MW extinction curve at low redshift ($z<1$, where $z$ is the redshift), supporting their
grain evolution model and their decomposition into separate species.
Because of the success, we further use the model of \citetalias{huang2021evolution} in this paper to predict another important observational property -- IR dust emission, as we explain in what follows.

The grain size distribution and grain composition predicted by simulations (especially \citetalias{huang2021evolution}) can also be constrained by the dust emission SED in the IR.
The wavelength dependence and features in the dust emission SED reflect the grain size distribution and dust composition in the following way \citep[e.g.][]{Desert90,draine2001infrared}.
Emission in the far-IR (FIR) region is contributed by large grains (typically $\sim 0.1~\micron$) which are
in radiative equilibrium with the ambient stellar radiation field.
In the mid-IR (MIR) (at wavelengths $\sim$ 10--60 $\micron$), the radiation is dominated by
very small grains, which are not in radiative equilibrium but
transiently heated by individual photons \citep[e.g.][]{Draine85}.
At near-IR (NIR) and MIR wavelengths, there are sharp features in the SED, especially those attributed to
polycyclic aromatic hydrocarbons (PAHs) \citep[e.g.][]{Leger84,Allamandola85}
or other carbonaceous species \citep{Sakata84,Duley93,Kwok11}. Because of these different contributions from different dust components,
\citet{Relano20} succeeded in extracting the mass ratio between small and large grains in nearby galaxies using SED fitting in the IR, and tested some of the above evolution models of grain size distribution in galaxies.

\citet[hereafter \citetalias{hirashita2020spectral}]{hirashita2020spectral} derived SEDs from the grain size distributions
calculated at different times in a one-zone galaxy evolution model. In their model, the dust is composed of silicate,
carbonaceous dust, and PAHs.
They broadly reproduced the MW SED and the IR colours of nearby galaxies.
Importantly, \citetalias{hirashita2020spectral} demonstrated that the MW SEDs may be explained as the result of continuous dust evolution subject to the changing conditions of the ISM over the lifetime of the galaxy.
However, their model treated the galaxy as an isolated, closed box; thus, whether or not the observed MW SED is consistent with theoretical expectations for the evolution of similar galaxies in the 
cosmological structure formation scenario is still an open question.

It is useful to determine whether the model of \citetalias{huang2021evolution}, which reproduces MW-like extinction curves for MW analogues in IllustrisTNG, also predicts dust emission SEDs similar to that of the MW. This will clarify whether the IR SED in the MW is consistent with cosmological expectations for the evolution of the Galctic ISM. There has been no study that explains the extinction curve and the IR SED in the MW simultaneously by solving for the full history of star formation, chemical enrichment and the grain size distribution consistently. Our aim in this paper is to provide such predictions for IR SEDs, based on the dust evolution model in \citetalias{huang2021evolution}. We concentrate on the dust emission SED in the IR, which we simply refer to as `the SED'.


This paper is organized as follows.
In Section~\ref{sec:model}, we explain the theoretical framework for the cosmological simulation, our dust evolution model and our SED calculations. In Section~\ref{sec:results}, we show the resulting SEDs for a sample of MW-like galaxies in the simulation. In Section~\ref{sec:discussion}, we discuss the results, focusing on parameter dependence and comparisons with other nearby galaxies. Section~\ref{sec:conclusions} provides our conclusions.
We adopt the solar metallicity $Z_{\sun}= 0.0127$ and the same cosmology as TNG: $h = 0.6774$, $\Omega_\Lambda= 0.6911$, $\Omega_\mathrm{m}= 0.3089$, and $\Omega_\mathrm{b}= 0.0486$ \citep{Planck:2016a}.

\section{Model}\label{sec:model}

We use the grain size distribution calculated by H21, who post-processed cosmological simulation data using dust evolution models. We briefly review these models and our method of deriving SEDs for a sample of simulated MW-like galaxies.

\subsection{MW-like galaxies in IllustrisTNG}\label{sec:sample}

H21 post-processed a cosmological simulation from the IllustrisTNG project \citep[TNG;][]{Pillepich18b, Springel18, Nelson18, Naiman18, Marinacci18}, a state-of-the-art cosmological galaxy formation simulation suite with a range of comoving volumes and resolution levels.
These simulations solve for the evolution of the dark matter and baryon density fields over cosmic history, and include comprehensive models of baryonic physics and energetic feedback from stars and active galactic nuclei \citep{Weinberger17,Pillepich18a}.
To obtain a sufficiently large sample of galaxies, we use the largest publically available simulation box with the coarsest mass resolution, TNG300-1, which has a side-length of 300 comoving megaparsecs \citep{Nelson19}.\footnote{\url{https://www.tng-project.org}} The available data include halo and subhalo catalogs, the physical properties of individual galaxies at multiple epochs, and galaxy and halo merger trees.

We use specific star formation rate (sSFR) and stellar mass ($M_*$) criteria to select our sample of MW-like galaxies. We do not use  morphological measurements (such as bulge to disk ratio) for the selection because their correspondence to observational measurements are less clear and they are less likely to be well converged at the resolution of TNG300. \citet{licquia2015improved} evaluated the global observational properties of the MW and derived a stellar mass $M_{*}=(6.08\pm{1.14})\times 10^{10}$ M$_{\sun}$ and sSFR ${}=(2.71\pm{0.59})\times10^{-11}$ yr$^{-1}$. We select 210 galaxies with stellar mass and sSFR in those 1-$\sigma$ ranges at $z=0$. H21 excluded four galaxies for which their model predicts unphysical dust-to-gas ratios likely due to mis-identification of subhalos. Thus, our final sample consists of 206 MW-like galaxies from TNG.

We trace the evolution of the following quantities along the merger tree of each galaxy in our sample:
the gas mass $M_\mathrm{gas}$, the stellar mass $M_*$, the SFR, the gas metallicity $Z$, and the silicate and carbon abundances ($Z_\mathrm{Si}$ and $Z_\mathrm{C}$, respectively). With these quantities, we calculate the evolution of the grain size distribution as summarized below (see \citetalias{huang2021evolution} for further details).

\subsection{Evolution of grain size distribution}
\label{sec:evolution}
Our dust evolution model (H21) is based on \citet{Hirashita20}, primarily formulated by \citet{Asano13} and \cite{Hirashita19}. It calculates the grain size distribution, denoted as $n(a,\, t)$, at each time-step. The grain size distribution is defined such that $n(a,\, t)\,\mathrm{d}a$ is the number density of grains with grain radii between $a$ and $a+\mathrm{d}a$. H21 further modified the model to incorporate the evolution of galaxy properties along the merger tree.
We consider each galaxy as a one-zone object, for which dust enrichment is calculated in a manner consistent with the metal enrichment. The total dust abundance is traced by the dust-to-gas ratio. In our calculation of the grain size distribution, we do not distinguish different dust species because of the difficulty in treating the interaction between multiple dust species.
Following \citetalias{huang2021evolution}, we adopt graphite properties in calculating the grain size distribution. Although the assumptions regarding the grain material affect the resulting grain size distribution, for example through material density, shattering and coagulation efficiencies \citep{Hirashita20}, a greater uncertainty is inherent in the one-zone treatment of the galaxies. We use the same one-zone setting and grain properties as \citetalias{huang2021evolution} (i.e.\ their fiducial model), which reasonably reproduced the MW extinction curve. For the purpose of calculating the SED, the calculated grain size distribution is decomposed into relevant species based on some quantities obtained from the simulation as explained later.

We consider five processes for dust evolution. Among them, stellar dust production and dust destruction in SN shocks both occur in the entire ISM. Dust enrichment by stellar dust production is traced by the metallicity increase given by TNG. For dust destruction by SN shocks, we adopt a grain-size-dependent destruction efficiency, which is multiplied by the sweeping rate of SN shocks to obtain the dust destruction rate. The other three processes take place only in particular ISM phases: grain growth through the accretion of gas-phase metals (which we simply call accretion) and coagulation occur in the dense ISM; grain disruption by shattering occurs predominantly in the diffuse ISM. 
For simplicity, we assume that the ISM is composed of dense and diffuse phases with  $({n}_\mathrm{H},\, {T}_\mathrm{gas}) = (0.3~\mathrm{cm}^{-3},\, 10^4~\mathrm{K})$ and $({n}_\mathrm{H},\, {T}_\mathrm{gas}) = (300~\mathrm{cm}^{-3},\, 100~\mathrm{K})$, respectively, where $n_\mathrm{H}$ is the hydrogen number density and $T_\mathrm{gas}$ is the gas temperature.
We calculate the dense-gas fraction($\eta_\mathrm{dense}$) using the following relation:
\begin{equation}
\textup{SFR}=\epsilon_{*}\frac{\eta_\mathrm{dense}{M^{*}_\mathrm{gas}}}{\tau_\mathrm{ff}},
\label{eq:eta_dense}
\end{equation}
where $M^{*}_\mathrm{gas}$ is the interstellar gas mass (the gas mass within twice the stellar half-mass radius), $\epsilon_{*}$ is the star-formation efficiency which we fix to 0.01, and $\tau_\mathrm{ff}=2.51$~Myr is the free-fall time evaluated for the dense ISM (i.e.\ at a density of $n_\mathrm{H}=300$~cm$^{-3}$).
The time evolution of the grain size distribution due to interstellar processing is also solved for:
the contributions from shattering and coagulation are solved in a single time-step $\Delta t$ with a weight of $(1-\eta_\mathrm{dense})$ and $\eta_\mathrm{dense}$, respectively. 
Accretion is also weighted for $\eta_\mathrm{dense}$.
For the parameters associated with shattering and coagulation, we adopt the fiducial model from H21 (see their section 3.1.4).
We discretize the entire grain radius range (3 \AA--10 $\micron$) into 64 logarithmic bins. We remove grains that evolve beyond the above radius range.

In H21, we extended \citet{Hirashita20}'s model to deal with growth of galaxies through smooth accretion and merging.
When the mass of gas associated with the galaxy increases, the metallicity changes (usually decreases because of dilution). Under the assumption that the accreted CGM contributes no dust and little metallicity to the galaxy, we scale the grain size distribution with metallicity to account for the dilution of metals.
Since TNG provides snapshot data with a time interval of roughly 0.1 Gyr, we do not know the exact time at which galaxies merge. We insert a virtual node before the descendant subhalo to estimate the gas mass of the two merging objects at the time of merging, evolve the grain size distributions to this virtual node, and take a weighted mean of $n(a,\, t)$ for the gas masses of the two merging subhalos to obtain the grain size distribution after the merger.

The overall evolution of grain size distribution
for the MW-like galaxies in the simulation is described as follows (see \citetalias{huang2021evolution} for details).
The grain size distribution is dominated
by large grains ($a\sim 0.1~\micron$) at $z\gtrsim 3$ because the dust production is dominated by stellar sources. Subsequently, the interplay between shattering and accretion rapidly increases the number of small grains ($a\lesssim 0.01~\micron$). The grain size distribution approaches that of \citetalias{Mathis77} at $z\lesssim 1$.
In the last phase, it is the balance between coagulation and shattering that drives the grain size distribution towards an \citetalias{Mathis77}-like distribution.

In the approach described above, we calculate the grain size distribution $n(a,\, t)$ for the sum of all the dust species. For the purpose of calculating extinction curves, \citetalias{huang2021evolution} decomposed the grain size distribution into relevant species based on \citet{Hirashita20}.
Following their method, we first separate the dust species into silicate and carbonaceous dust according to the abundance ratio of Si to C given by TNG for each subhalo at each epoch. The carbonaceous component is further divided into aromatic and non-aromatic species with ratio $(1-\eta_\mathrm{dense}) : \eta_\mathrm{dense}$, noting that aromatization predominantly occurs in the diffuse medium. The aromatization time-scale is so short that the aromatic fraction can be approximated with the equilibrium value $1-\eta_\mathrm{dense}$ \citep{Rau19,Hirashita20}.
We neglect photodestruction of PAHs by locally intense UV irradiation \citep[e.g.][]{Madden:2006a,Khramtsova:2013a,Murga:2019a} since we are interested in the diffuse radiation field for our one-zone model.

\subsection{Calculation of SED}\label{subsec:SEDmodel}

We calculate the SED based on the grain composition and the grain size distribution following \citetalias{hirashita2020spectral} (except for the treatment of PAH ionization) based on the framework developed by \cite{li2001infrared} and \cite{draine2001infrared}.
We now briefly review our method for calculating  SEDs.

We assume that the IR radiation is optically thin. We calculate the intensity at frequency $\nu$ per hydrogen, $I_\nu$ as
\begin{equation}
I_{\nu}=\sum_{i}\int_{0}^{\infty}\mathrm{d}a\frac{1}{n_\mathrm{H}}n_{i}(a)\upi a^2 Q_\mathrm{abs}(a,\,\nu)\int_{0}^{\infty}\mathrm{d}T B_{\nu}(T)\frac{dP_{i}}{dT},
\label{eq:sed}
\end{equation}
where $Q_\mathrm{abs}(a,\nu)$ is the absorption cross-section normalized to the geometric cross-section, $n_i$ is the grain size distribution with $i$ indicating the grain species, $B_\nu (T)$ is the Planck function at frequency $\nu$ and temperature $T$, and $\mathrm{d}P_i/\mathrm{d}T$ is the temperature distribution function of a grain with radius $a$.
Note that our post-processing dust model outputs the grain size distribution per gas mass, or $n(a)/n_\mathrm{H}$; thus, we output SED per hydrogen.
The temperature distribution function is obtained based on the balance between heating by the interstellar radiation field (ISRF) and radiative cooling. The thermal emission from the dust is weighted by the obtained probability distribution function of dust temperature.
In calculating the dust heating, we adopt the ISRF spectrum appropriate for the MW from \cite{mathis1983interstellar}.
Indeed, the MW analogs we select from TNG shows a narrow range of SFR (1--2 M$_{\sun}$ yr$^{-1}$);
thus, it would be reasonable to assume that the ISRFs of the sample are all similar to that of the MW.
We also separately examine the effect of ISRF intensity
(the ISRF is scaled by the parameter $U$; i.e., $U=1$ corresponds to the standard case) in Section~\ref{subsec:U}.

We decompose the grain size distribution into silicate and carbonaceous dust (aromatic and non-aromatic) as explained in Section \ref{sec:evolution}. Following \citetalias{hirashita2020spectral}, we adopt the properties (optical properties and specific heat capacity) from \citet{draine2007infrared}.
The ionization of PAHs depends on the detailed density
structures of the ISM, which are different from galaxy to galaxy. Thus,
for simplicity, we neglect the ionization of PAHs (i.e.\ we adopt the optical properties of neutral PAHs by default), but we separately examine the effect of PAH ionization using the optical properties of ionized PAHs in Section \ref{subsec:ionization}. The realistic SED should lie between these two cases.

Fixing $U$ and the PAH optical parameters is useful for the purpose of concentrating on the variations caused by the grain size distribution and composition.

\section{Results}\label{sec:results}

In this section, we present the SEDs we derive for our MW-like sample.
Since we have already present the evolution of the grain size distribution in these galaxies in H21, we focus on their SEDs at $z=0$, in order to examine whether the calculated grain properties reproduce the observed MW SED.

\subsection{SEDs of the entire MW-like sample}\label{subsec:SED_all}

\begin{figure}
    \includegraphics[width=0.5\textwidth]{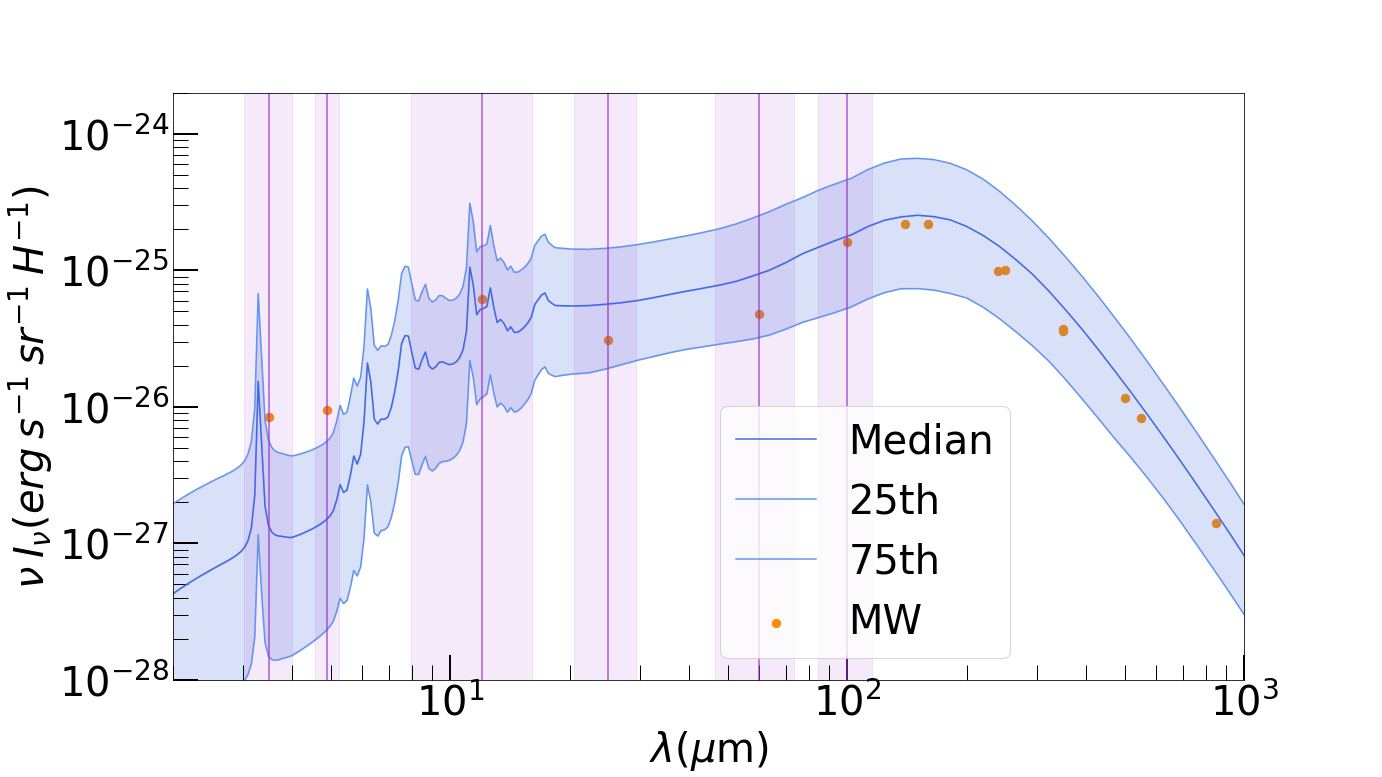}
    \caption{SEDs of MW-like galaxies in TNG at $z=0$. SEDs are shown by the intensity per hydrogen multiplied by the frequency $\nu I_\nu$ (also in the following figures). The solid line traces the median at each wavelength. The shaded region bounded by the thin lines presents the 25th and 75th percentile range of those SEDs. The circles show the observational MW data from \citet{compiegne2011global}.  The errors on these data points are much smaller than the dispersion of the theoretical SEDs. The purple bands correspond to the width of the band-pass for each observational datapoint at $\lambda\leq 100~\micron$. We do not show the band-passes at longer wavelength, since the bands are heavily overlapped.}
    \label{fig:galaxy}
\end{figure}

We show the statistical result for the SEDs of the entire MW-like sample in Fig.\ \ref{fig:galaxy}. As mentioned in Section \ref{subsec:SEDmodel}, we show the intensity per hydrogen for the SED. We present the median and 25th and 75th percentiles at each wavelength. We compare the result with the observed MW SED, for which we take the data for the diffuse high Galactic latitude medium from \citet{compiegne2011global}. These data can be treated as typical of dust emission in the Solar neighbourhood \citep[see also][]{Desert90}. The same data set was also used in \citetalias{hirashita2020spectral}.

Considering the dispersion of our results,
the observational data are broadly consistent with the SEDs
we obtain for our TNG MW sample.
In particular, the SED at $\gtrsim 100~\micron$ is consistent with the median, which means that we correctly reproduce the dust-to-gas ratio (or dust abundance relative to hydrogen).
The observed intensities at 20--60 $\micron$ are well within the dispersion of our predictions.
There are, however, some discrepancies in detail
seen in the two shortest-wavelength bands ($<10~\micron$), where PAHs dominate the emission.
An underestimate of PAH emission is also observed in our previous one-zone model by \citetalias{hirashita2020spectral}, who argued that
the discrepancies may be due to the treatment of the dense gas fraction ($\eta_\mathrm{dense}$). Lowering $\eta_\mathrm{dense}$ would increase the PAH emission because the aromatic fraction increases.
However, lowering $\eta_\mathrm{dense}$ would also lead to enhanced shattering, which increases the overall small-grain abundance. This would further enhance the emission at MIR wavelengths ($\sim 20$--60 $\micron$) and thereby increase the difference between the median SED and the observational data. This implies that the underestimate of PAH emission needs to be resolved in a way that does not break the agreement at MIR wavelengths. We discuss some possible solutions to the discrepancy at PAH-dominated wavelengths in Section \ref{subsec:improvements}.

Although our median SED does not perfectly match the observed MW SED, our model can still be considered broadly successful if we consider the dispersion and the uncertainties in the model.
To further discuss the dominant source of the dispersion in our SEDs, we now examine the effects of dense gas fraction and metallicity.

\subsection{Dependence on galaxy properties}

\citetalias{huang2021evolution} focused on four galaxy properties: SFR, stellar mass, gas mass, and metallicity. The dense gas fraction calculated by equation (\ref{eq:eta_dense}) is also important in determining the balance between dust growth (accretion and coagulation) and dust disruption (shattering). The distributions and evolution of these parameters are given in \citetalias{huang2021evolution}. Here we examine the impact of these parameters on our SED predictions.

\begin{figure}
    \includegraphics[width=0.5\textwidth]{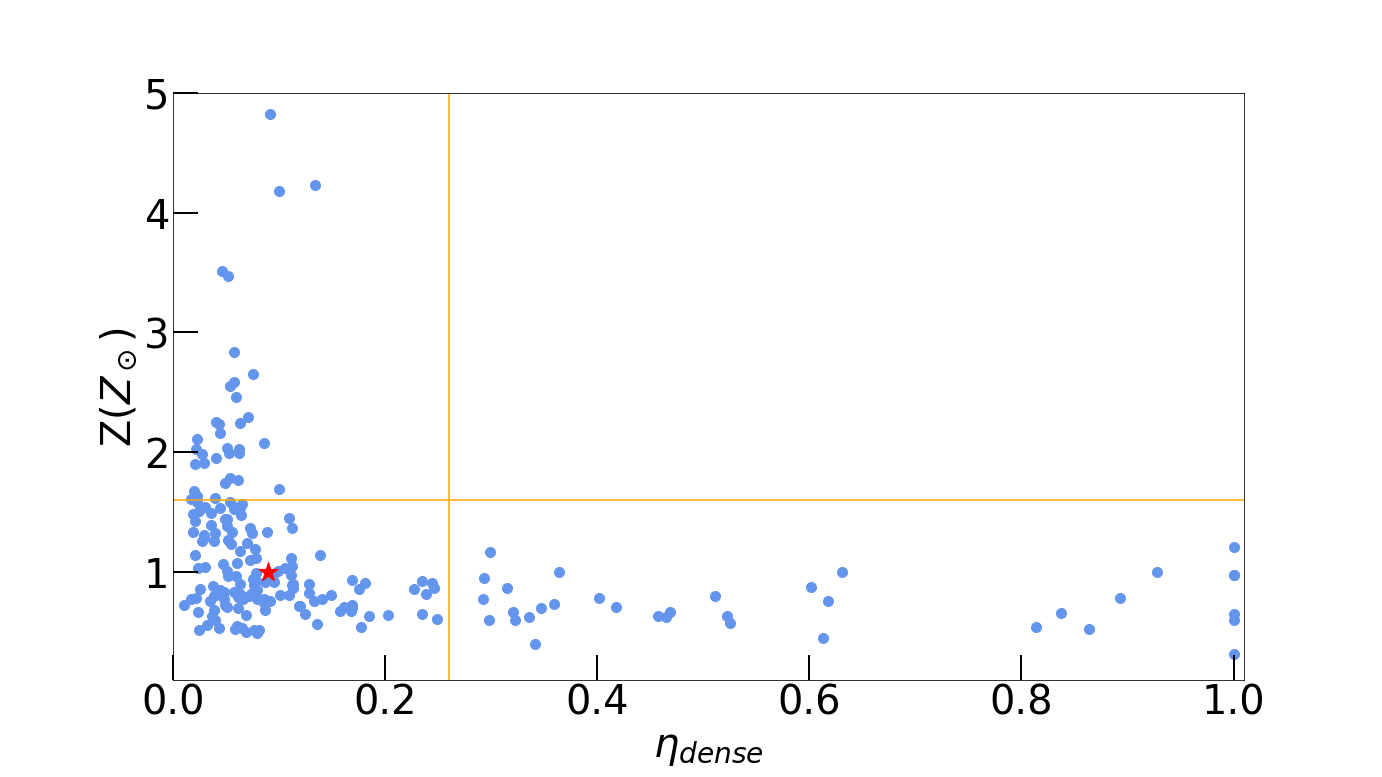}
    \caption{Relationship between dense gas fraction $\eta_\mathrm{dense}$ and gas metallicity $Z$ in our sample of simulated MW-like galaxies. The orange lines at $Z=1.6{{Z}_\odot}$ and $\eta_\mathrm{dense}=0.26$ are the thresholds used to divide our sample into high/low-$Z$ and high/low-$\eta_\mathrm{dense}$ as described in the text. The red star marks the estimated values for the real MW (see text).}
    \label{fig:DFtoZ}
\end{figure}

Among the above quantities, the metallicity and the dense gas fraction have significant correlation with the SED. We also found that there is a strong correlation between the metallicity and the dense gas fraction.
In Fig.\ \ref{fig:DFtoZ}, we show the relation between these two quantities for our MW-like sample. High-metallicity objects with $Z>1.6$ Z$_{\sun}$ are only found at low $\eta_\mathrm{dense}$, while galaxies around Solar metallicity have a wide range of $\eta_\mathrm{dense}$. Note that the dense gas fraction is estimated using equation (\ref{eq:eta_dense}) and hence is a function of the star formation efficiency, $\mathrm{SFR}/M_\mathrm{gas}^*$.
Thus, low/high $\eta_\mathrm{dense}$ is equivalent to a low/high fraction of gas associated
with star formation.
The trend in Fig.\ \ref{fig:DFtoZ} indicates that high-metallicity galaxies have less fractions of gas directly associated with star formation. It is probable that those galaxies already consumed dense gas and caused more chemical enrichment (reflected in their high metallcities).

For reference, we estimate $\eta_\mathrm{dense}$ according to the above definition based on the gas mass and SFR observed in the MW. We use $M_\mathrm{gas}^*\sim 5\times 10^9$ M$_{\sun}$ \citep{Mathis00} and
SFR $\sim 1.7$ M$_{\sun}$ yr$^{-1}$ \citep{licquia2015improved}, respectively, and thus obtain
$\eta_\mathrm{dense}=0.09$ from equation (\ref{eq:eta_dense}).
Although both quantities have large uncertainty, Fig.\ \ref{fig:DFtoZ} shows that, taking $Z=1$ Z$_{\sun}$, our MW estimate is consistent with the distribution of simulated galaxies.
Indeed, the median values of our MW-like sample in the simulation are $Z=0.9$ Z$_{\sun}$ and $\eta_\mathrm{dense}=0.07$, which are in good agreement with the above.

Based on the distribution in Fig.\ \ref{fig:DFtoZ}, we divide the sample into high and low metallicity subsets at $Z=1.6$ Z$_{\sun}$, and into high and low dense-gas fraction subsets at $\eta_\mathrm{dense}=0.26$. These divisions are marked in Fig.\ \ref{fig:DFtoZ}.
We find that there is no significant dependence of the SEDs on other parameters, so we concentrate on the dependence on the metallicity and the dense gas fraction below.

\subsubsection{Dependence on dense gas fraction}\label{subsec:eta_dense}

\begin{figure}
    \includegraphics[width=0.5\textwidth]{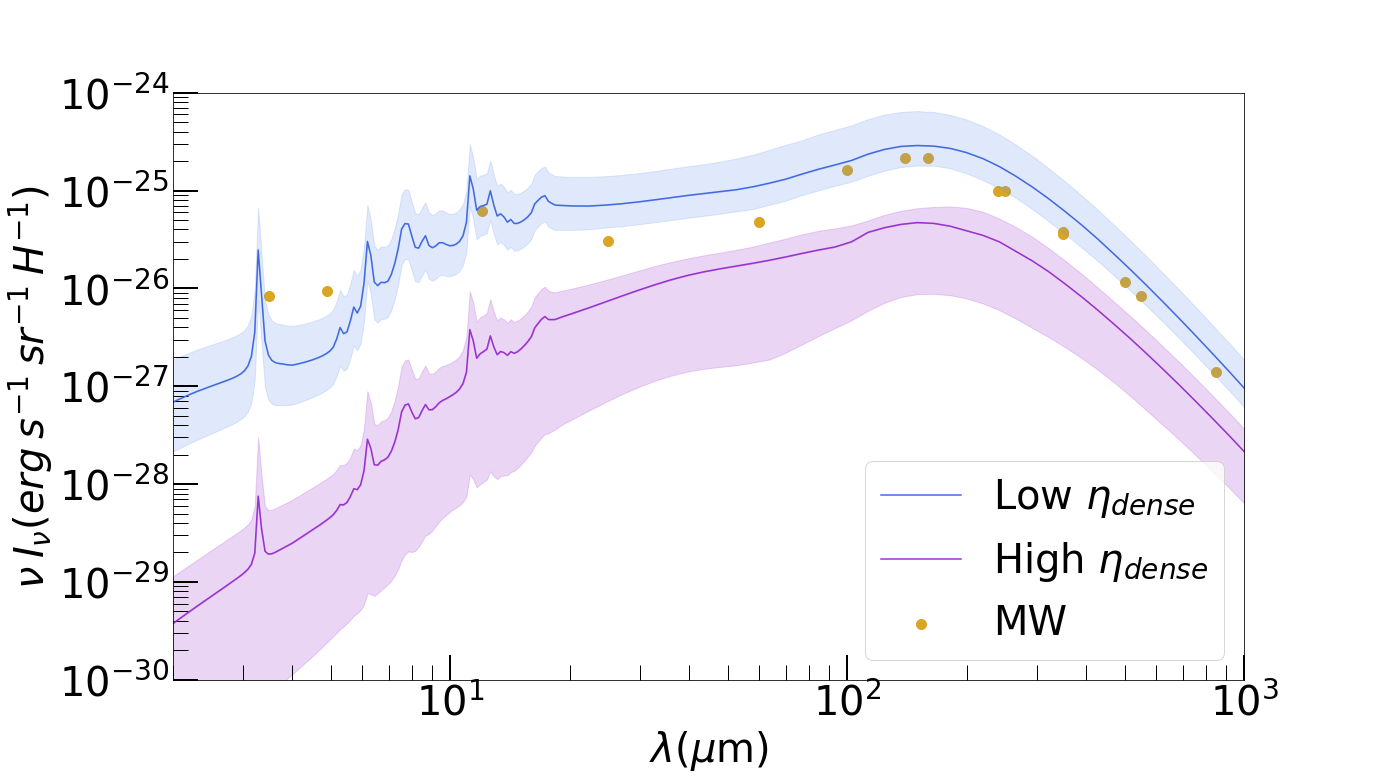}
    \includegraphics[width=0.5\textwidth]{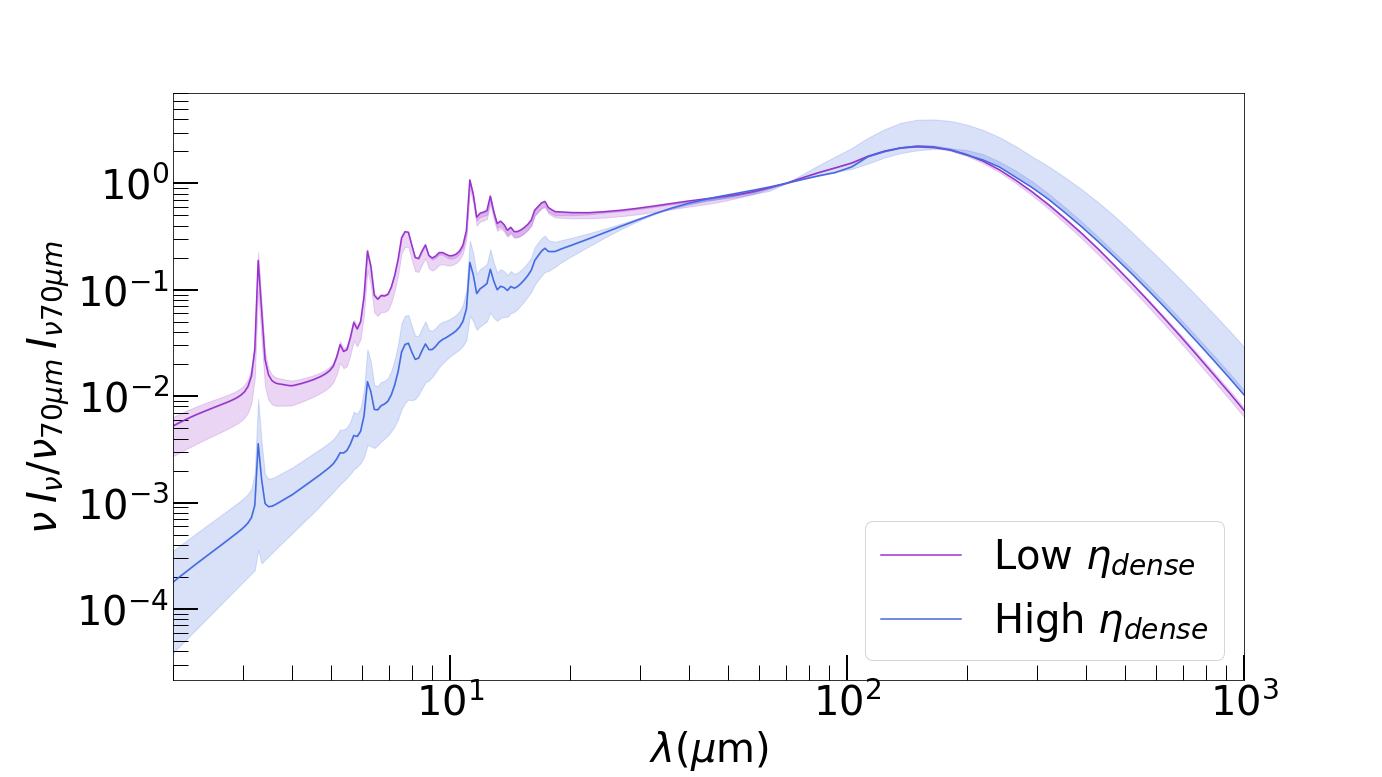}
    \caption{Upper panel: SEDs of our subsamples with high (purple) and low (blue) dense gas fraction, divided at $\eta_\mathrm{dense}=0.26$. Lines correspond to the median SED and shaded regions show the 25th to 75th percentile range.
    Lower panel: The same SEDs normalized to the intensity at $\lambda =70~\micron$, highlighting the difference in SED shape.}
    
    \label{fig:DFSED}
\end{figure}

The dense gas fraction is one of the most important parameters in our model, because it determines the balance between dust growth and dust disruption, as well as the aromatic fraction (Section \ref{sec:evolution}). We show the SEDs of the high and low-$\eta_\mathrm{dense}$ subsamples in Fig.\ \ref{fig:DFSED} (upper). We observe that the high-$\eta_\mathrm{dense}$ subsample has significantly lower intensities at all wavelengths. Overall, this reflects the difference in dust abundance: the lower-$\eta_\mathrm{dense}$ subsample has a higher dust abundance because it has higher metallicity, as is clear in Fig.\ \ref{fig:DFtoZ}. Moreover, we observe that the difference between the two subsamples is not uniform in wavelength; at short wavelengths the SEDs (including the PAH features) are enhanced in the lower-$\eta_\mathrm{dense}$ sub-sample.
This is consistent with the results of \citetalias{hirashita2020spectral}: lower $\eta_\mathrm{dense}$ favours PAH emission because of more small-grain production and more aromatization.

To clarify the difference in SED shape, we show the SEDs normalized to the intensity at 70~$\micron$ in Fig.~\ref{fig:DFSED} (lower panel; the reference wavelength is not important as long as it is in the FIR regime).
We find that the difference between the two subsamples is larger at short wavelengths ($\lesssim 20~\micron$). This indicates that the two subsamples have not only different dust abundances but also different grain size distributions. Indeed, we confirmed that the there is an excess of submicron grains in the high-$\eta_\mathrm{dense}$ sample. Recall that a high dense gas fraction leads to an enhancement of coagulation and a suppression of shattering in our model. The aromatic fraction is also lower in the case of higher $\eta_\mathrm{dense}$ in our model, leading to weaker PAH features.


\subsubsection{Dependence on metallicity}\label{subsec:z}

\begin{figure}
    \includegraphics[width=0.5\textwidth]{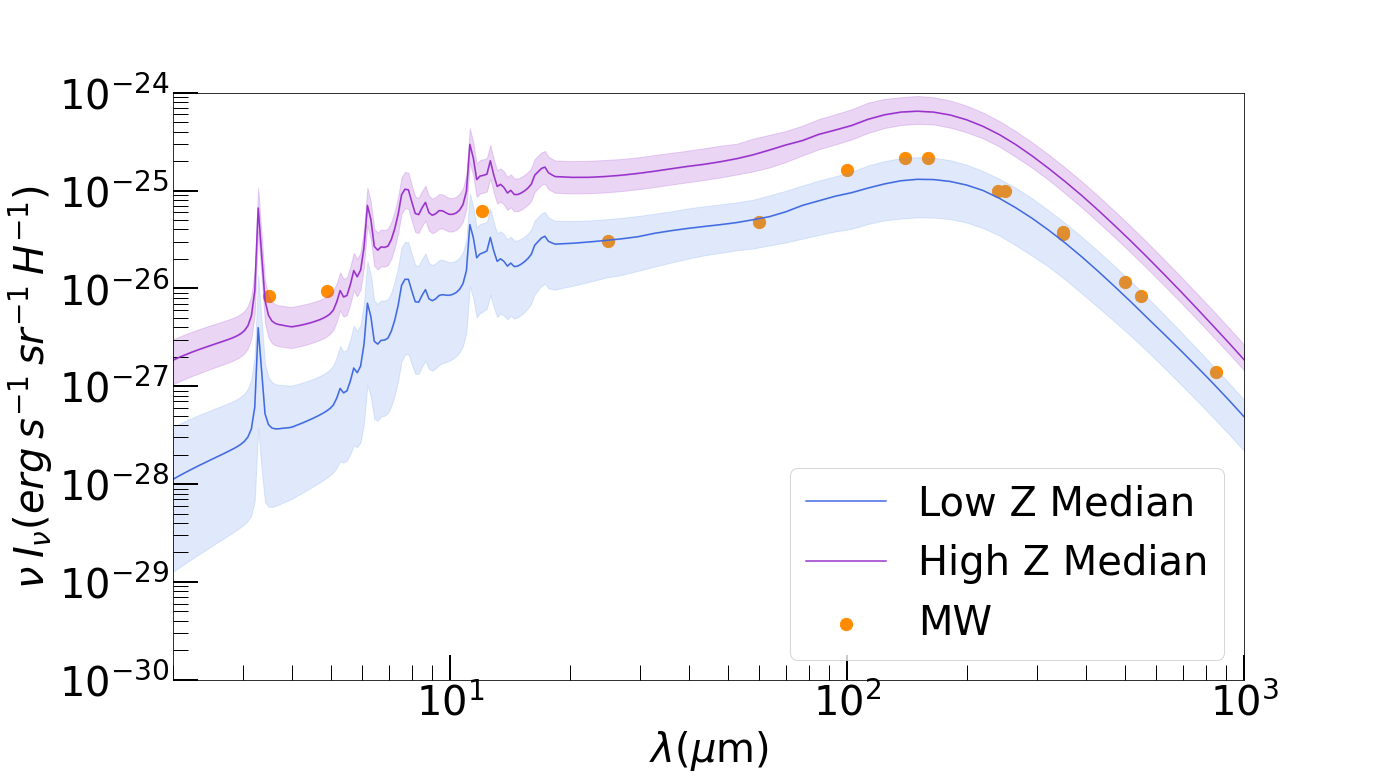}
    \includegraphics[width=0.5\textwidth]{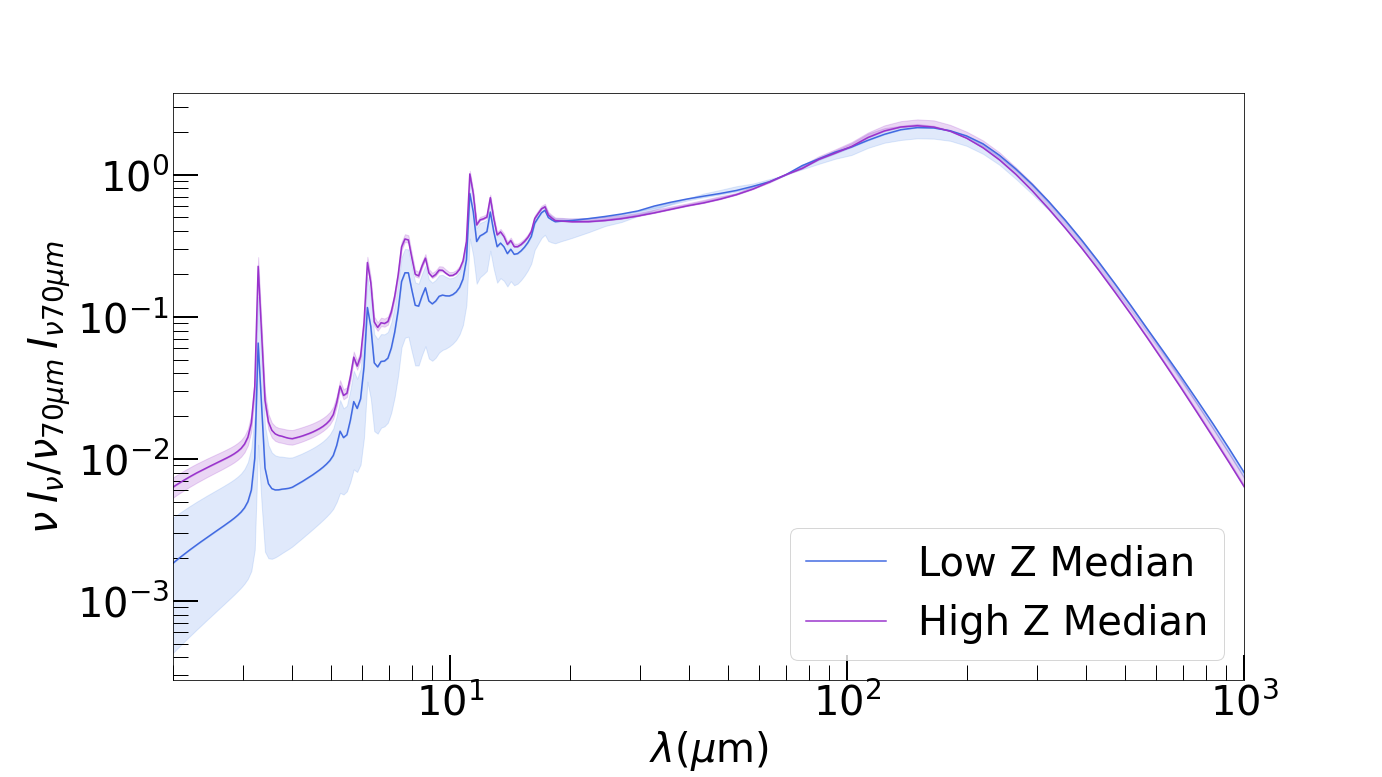}
    \caption{Upper panel: SEDs of our high- (purple) and low-metallicity (blue) subsamples, divided at $Z=1.6$~Z$_{\sun}$. Lines correspond to the median SED and shaded regions show the 25th to 75th percentile range. Lower panel: the same SEDs normalized to the intensity at $\lambda =70~\micron$, highlighting the difference in SED shape.
    }
    \label{fig:ZSED}
\end{figure}

We compare the SEDs of the low- and high-metallicity subsamples in Fig.\ \ref{fig:ZSED}.
Recall that the two subsamples are divided at $Z=1.6$ Z$_{\sun}$.
As expected, the metallicity mainly affects the overall intensity because higher metallicity implies higher dust abundance. As shown by \citetalias{huang2021evolution} (their fig.\ 7), there is a strong correlation between the metallicity and the dust-to-gas ratio. 
This correlaion is also supported by observational data of nearby galaxies \citep[e.g.][]{Remy-Ruyer14}.
The dispersion in the dust-to-gas ratio is larger at lower metallicity (\citetalias{huang2021evolution}); this is why we find a larger dispersion in the SEDs of the lower-$Z$ subsample. The dispersion in the dust-to-gas ratio is mainly caused by different efficiencies of dust growth (accretion), which is the major dust formation mechanism at metallicities $\gtrsim 0.3$ Z$_{\sun}$ (H21; see also \citealt{DeVis21}). Since dust growth is saturated at high metalllicity, a large fraction of metals are locked into dust in the high-$Z$ subsample, while a larger variation in the dust-to-gas ratio appears in the low-$Z$ subsample.

As mentioned above, $Z$ is also correlated with $\eta_\mathrm{dense}$. Higher $\eta_\mathrm{dense}$ leads to an enhancement of large grains relative to small grains (through coagulation) and comparatively suppressed emission at short wavelengths. This is confirmed by the comparison of the SEDs normalized to the value at 70 $\micron$ for the two subsamples in Fig.\ \ref{fig:ZSED} (lower panel).
Since the main effect of metallicity is to scale the total dust abundance, the difference in the SED shape is smaller between the metallicity subsamples than between the $\eta_\mathrm{dense}$ subsamples. The dispersion at short wavelengths, where PAH emission dominates, is large in the low-$Z$ subsample, showing that the grain size distribution also varies more at low $Z$.

\subsection{Redshift evolution}

\begin{figure}
    \includegraphics[width=0.5\textwidth]{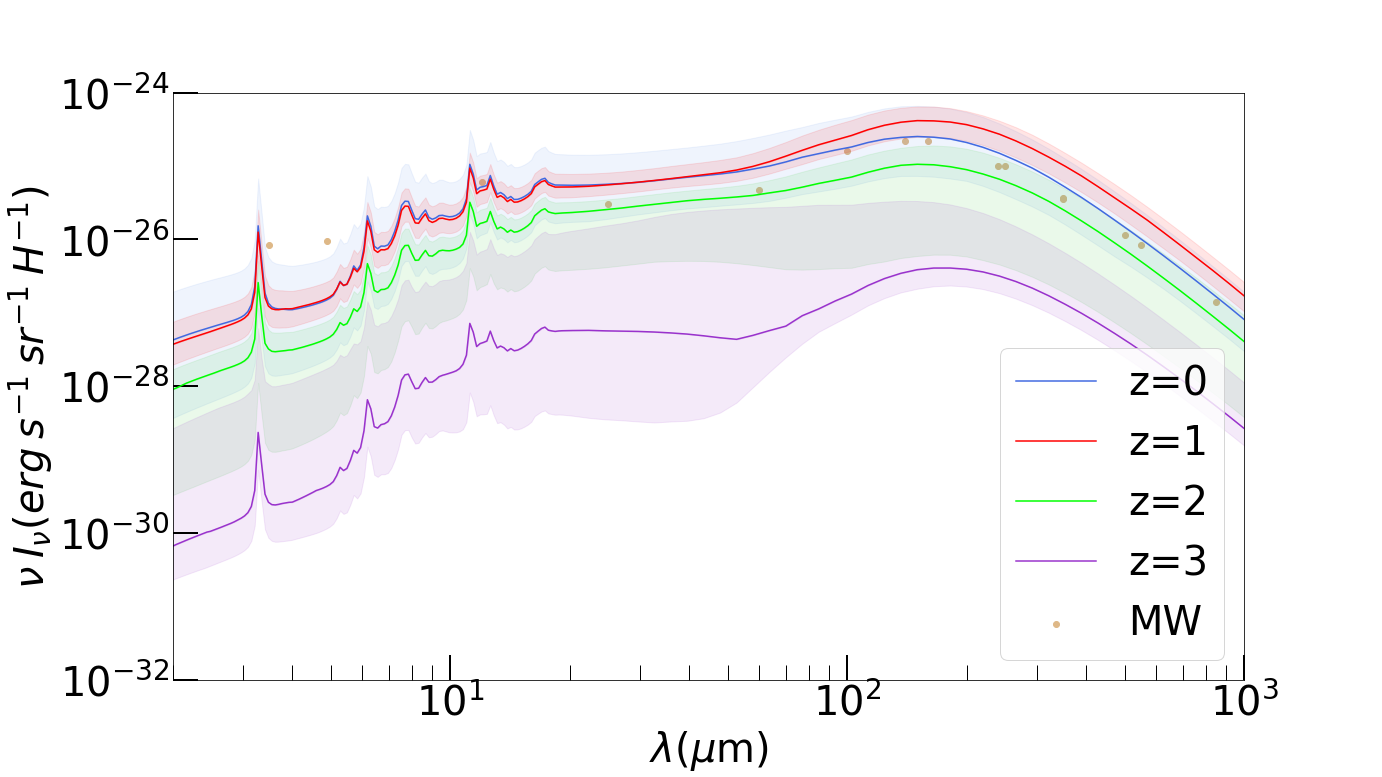}
    \caption{SEDs of the main-branch progenitors for the TNG MW analogs at $z=1$--$3$ together with their SEDs at $z=0$. The purple, green, red, and blue solid lines show the medians at $z=3$, 2, 1, and 0, respectively.
    Shaded areas with the same colours indicate the 25th to 75th percentile range. The dots are the same observational data for the MW SED as in the above figures.}
    \label{fig:z_evo_SED}
\end{figure}

Although our purpose is to compare our simulated SEDs to that of the MW at $z=0$, it is useful to examine the SEDs of their progenitors. This serves to clarify the SED evolution of MW-like galaxies. \citetalias{huang2021evolution} already discussed the evolution of the grain size distributions and extinction curves along the main branch (tracing the largest progenitors) of the merger trees of each galaxy at $z=1$, 2, and 3.
In Fig.\ \ref{fig:z_evo_SED} we present the SEDs of the main-branch progenitors at $z=1$--$3$.

We observe in Fig.\ \ref{fig:z_evo_SED} that, overall, there is a trend of increasing intensity from high to low redshift, which reflects the dust enrichment.
The median simulated SED approaches a level consistent with the $z=0$ MW SED at $z\sim 1$. At $z\sim 1$ the typical FIR intensity exceeds that at $z\sim 0$, reflecting an excess of dust abundance at $z\sim 1$: \citetalias{huang2021evolution} interpreted this as a dilution of dust abundance by gas accretion from $z\sim 1$ to $z\sim 0$.
At $z\sim 3$, the intensity is significantly lower than that at $z\sim 0$. Thus, the `growth' of dust emission occurs after $z\sim 3$. At $z\sim 3$, the SEDs not only have lower intensities but also a larger dispersion. In particular, the dispersion is largest in the MIR wavelength, reflecting a large range in the abundance of small grains. \citetalias{huang2021evolution} showed that most of the grain size distributions at $z\sim 3$ have a bump at $a\lesssim 0.01~\micron$, created by the growth (through accretion) of shattered grains. Since the efficiency of this process is sensitive to the dust and metal abundances and the dense gas fraction, the variation of these quantities results in a large scatter in the abundance of small ($a\lesssim 0.01~\micron$) grains, which contribute to the MIR emission.
Thus, the redshift evolution shown here implies that the SEDs of $z=0$ MW-like galaxies are more diverse at higher redshift.

The evolution of PAH emission at 8~$\micron$ is traced by the \textit{Spitzer} 24~$\micron$ band
at $z\sim 2$. \citet{Shivaei:2017a} showed, based on stacked data, that the PAH emission strength correlates with the metallicity at $z\sim 2$.
This redshift may be an interesting epoch for the buildup of PAH emission. Our result indicates that the PAH emission of MW progenitors reaches a level as strong as that in the present-day MW
at $z\sim 2$ (Fig.\ \ref{fig:z_evo_SED}). Since we predict a large evolutionary gap in the intensity level
between $z\sim 2$ and $3$, it will be interesting to sample PAH emission at $z\sim 3$ in the future.

\subsection{Comparison with nearby galaxies}

Although we have concentrated on the MW, we expect that our model is applicable to other galaxies similar to the MW.
Among the samples used by \citetalias{hirashita2020spectral}, we adopt the one from \citet{engelbracht2008metallicity},
which has more objects around Solar metallicity than the other sample
taken from \citet{Dale:2017a}.
The results and discussions below do not change if we include the data from \citet{Dale:2017a}.
For the observational sample, we adopt $12+\log (\mathrm{O/H})_{\sun}=8.69$
for the Solar oxygen abundance
\citep{Lodders:2003a} and assume that the metallicity scales with the oxygen abundance.
The typical errors in  luminosity and  metallicity are 10 per cent and 0.1~dex, respectively.
We use only  galaxies with metallicity comparable to the MW, defined by
$Z>0.5$~Z$_{\sun}$ [$12+\log (\mathrm{O/H})_{\sun}>8.39$]. This metallicity range is  consistent
with our selection of MW-like galaxies in the simulation (\citetalias{huang2021evolution}).
This criterion selects 16 galaxies from \citet{engelbracht2008metallicity} for comparison
to our simulation results.
For each galaxy, we use the stellar-subtracted fluxes at 3.6, 4.5, 5.8, 8, 24, 70, and 160~$\micron$.
To cancel out the scaling of the SED with the total dust mass,
we divide the flux in each band by that at 70~$\micron$, and concentrate on the SED shape.
We also normalize our simulated SEDs to the value at 70 $\micron$. This normalization is  convenient to check the 160 $\micron$--70 $\micron$ colour, which is a good indicator of the equilibrium dust temperature determined by the ISFR ($U$).

\begin{figure}
    \includegraphics[width=0.5\textwidth]{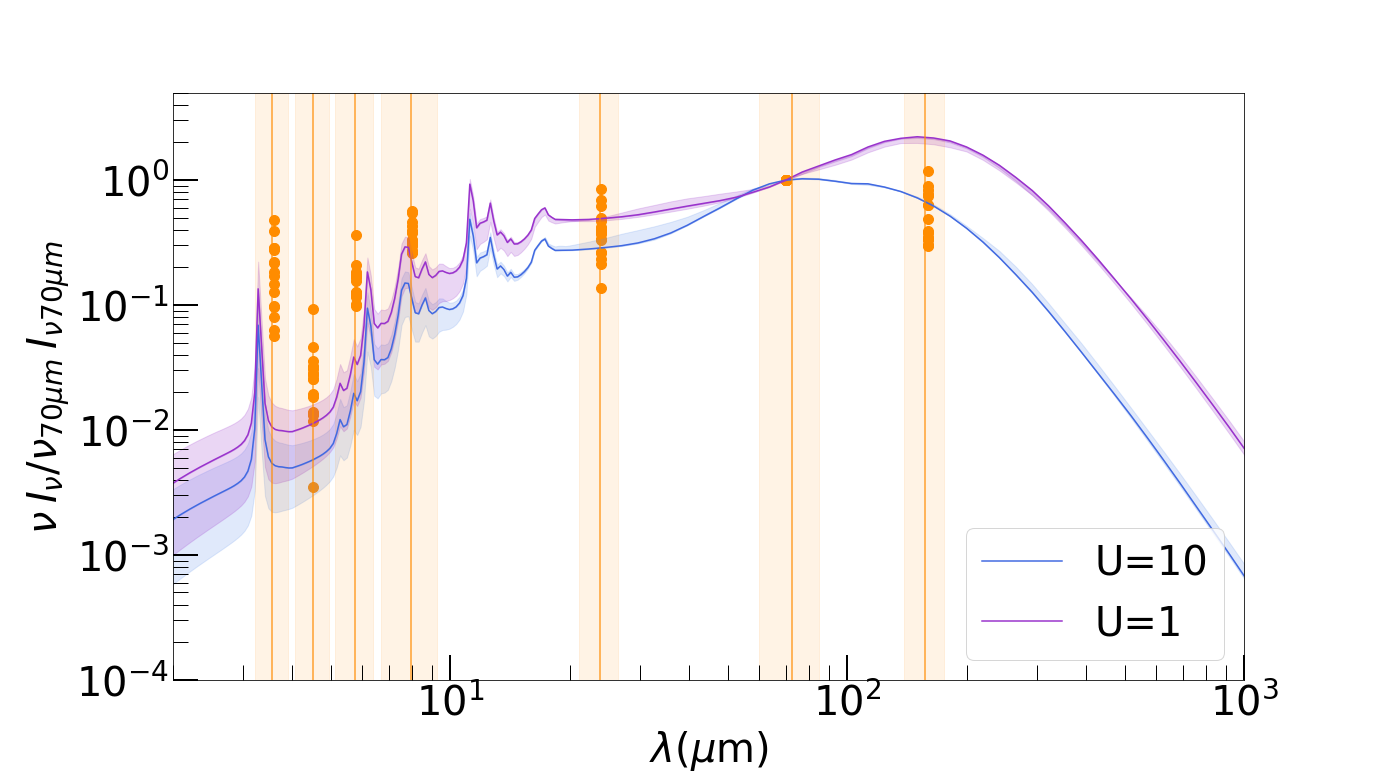}
    \caption{Normalized SEDs for $U=1$ and $U=10$, compared to observational data for nearby galaxies. All of the SEDs and data points are normalized to the intensity at 70 $\micron$. The blue and purple lines present results for $U=10$ and $U=1$, respectively. The dots show the data of 16 galaxies from \citet{engelbracht2008metallicity} at 3.6, 4.5, 5.8, 8, 24, 70 and 160 $\micron$ (note that we only have a single point at 70 $\micron$ because this wavelength is adopted for normalization). Shaded areas with the same colour indicate the 25th to 75th percentile ranges. The vertical bands associated with the observational data represent the width of the bandpass.}
    \label{fig:nearby}
\end{figure}

Fig.\ \ref{fig:nearby} shows the results of this comparison.
To take into account the variation in the FIR colours
(160 $\micron$--70 $\micron$ flux ratios), we plot our results for $U=10$ in addition to our default choice of $U=1$ (the MW ISRF).
We find, however, that the SED shapes at $\lambda <70~\micron$ are not sensitive to $U$. This is because the SEDs at such short wavelengths are dominated by stochastic heating of dust grains and simply scale with $U$, without changing the overall SED shape
\citep{draine2007infrared}. We observe that the SED shape, normalized at 70 $\micron$, is consistent with the observations at 24 $\micron$, but tends to underpredict the normalized intensities at $\lambda\leq 8~\micron$.
This underestimate is also seen for the MW SED (Section \ref{subsec:SED_all}).
Possible improvements are discussed in Section \ref{subsec:improvements}.
We emphasize that our models nevertheless predict the lower part of the observed SEDs well, even at the PAH-dominated wavelengths.

\section{Discussion}\label{sec:discussion}

In this section, we further examine the dependence of our results on  parameters that we held fixed in the analysis above,
in order to clarify factors that may influence the resulting SEDs.
We focus on parameters that affect the intensity level directly, that is, the PAH ionization and the ISRF intensity $U$.

\subsection{Effects of ionized PAHs}\label{subsec:ionization}

In Section \ref{sec:results}, we assumed neutral PAHs for simplicity, because the PAH ionization state depends on the physical quantities of the ambient medium, which is not resolved in our calculation. Here we examine the effect of PAH ionization on the SEDs.

\begin{figure}
    \includegraphics[width=0.5\textwidth]{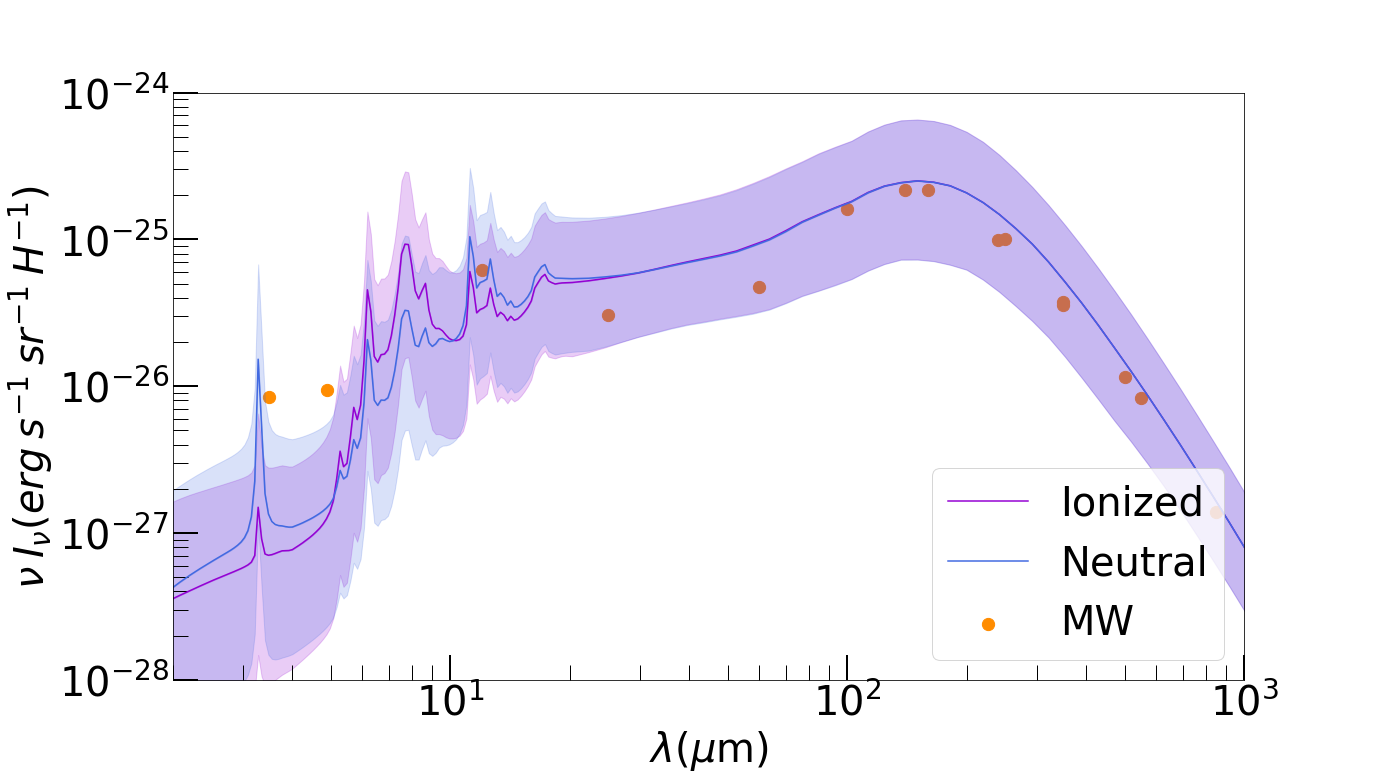}
    \caption{SEDs assuming ionized (purple) and neutral (blue) PAHs. Lines show the sample medians and shaded areas the 25th to 75th percentile ranges. Note that we use a narrower range for the vertical axis than in previous figures (e.g.\ Fig.\ \ref{fig:galaxy}) and that the two cases overlap almost completely at $\lambda\gtrsim 20~\micron$. The dots are the MW data also shown previously. 
    }
    \label{fig:ion}
\end{figure}

In Fig.\ \ref{fig:ion}, we compare the SEDs calculated using neutral and ionized PAHs.
We observe that the ionized PAHs show weaker 3.3 and 12 $\micron$ features but stronger emission around 8 $\micron$. Therefore, the change of the PAH emission caused by PAH ionization is not monotonic with wavelength. This means that the ionization of PAHs does not explain the systematic underprediction of the median SED at $\lambda\lesssim 10~\micron$. We also note that the dispersion of the SEDs in the sample is larger than the difference between the ionized and neutral PAHs. We conclude that the effect of PAH ionization is subdominant, compared to the expected scatter in the SEDs among MW-like galaxies.

\subsection{Variation of ISRF intensity}\label{subsec:U}

In our main analysis we fixed the ISRF intensity $U$ (Section \ref{subsec:SEDmodel}).
Nevertheless, our sample has a factor two dispersion in SFR.
Thus, it is still worth examining how much the SEDs are affected by a change in the ISRF; in particular, it is interesting to examine whether the systematic underpredicton at $\lambda\lesssim 10~\micron$ can be improved by increasing $U$. Based on the variation in the SFR, we change the ISRF by a factor of 2; that is, we compare the SEDs with $U=1$ and $U=2$.

\begin{figure}
    \includegraphics[width=0.5\textwidth]{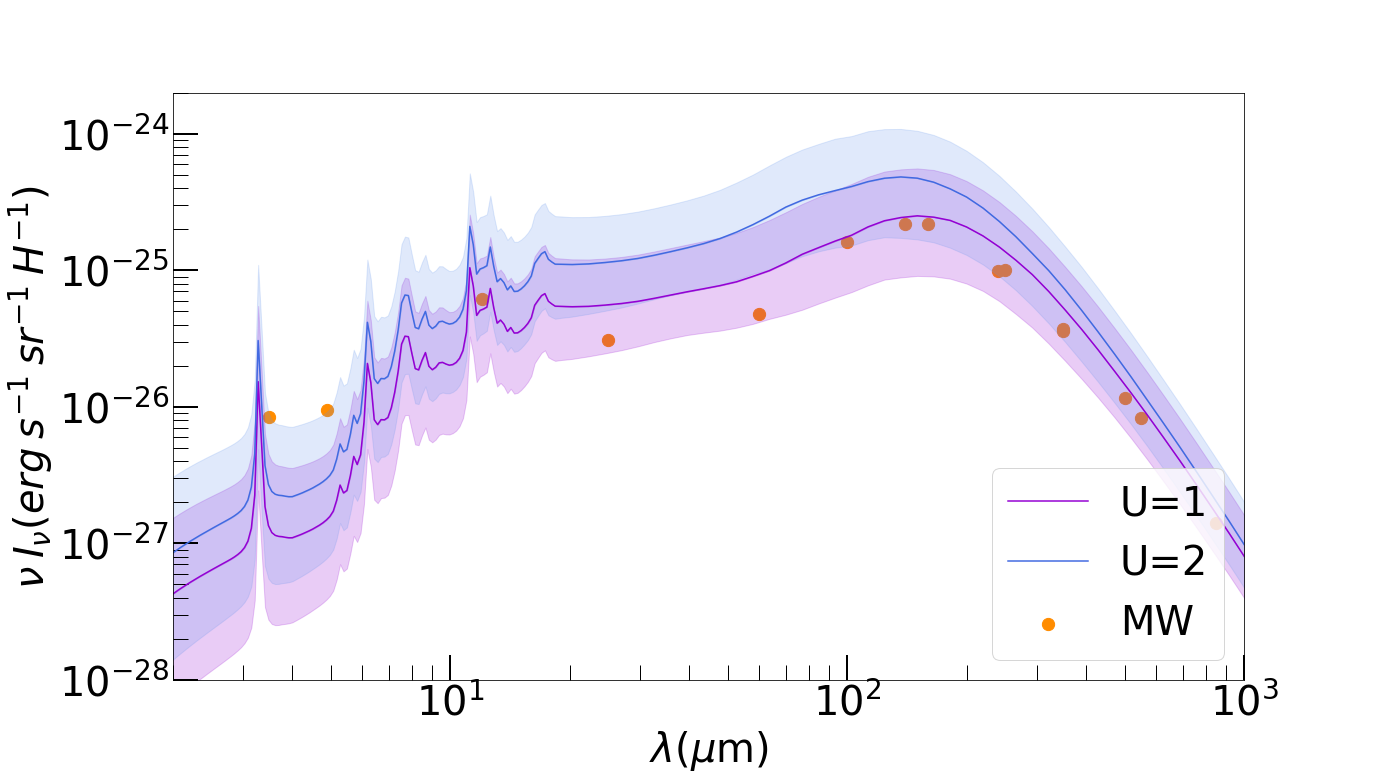}
    \caption{SEDs assuming $U=2$ (blue) compared with those with $U=1$ (purple), for our entire MW-like galaxy sample. Lines show medians and the shaded area with the same colour the 25th to 75th percentile ranges.
    The dots are the MW data also shown previously.}
    \label{fig:USED}
\end{figure}

In Fig.\ \ref{fig:USED}, we compare the SEDs for $U=1$ and 2.
As expected, the intensity is higher at all wavelengths for higher $U$.
In particular, the prediction with $U=2$ improves the fit to the MW data at $\lambda\lesssim 10~\micron$.
However, it significantly overpredicts the data points at longer wavelengths.
This confirms that the standard MW radiation field $U=1$ is more consistent with the MIR--FIR SED and that increasing $U$ is not the solution for an overall better fit to the data.

\subsection{Possible improvements and alternative models}\label{subsec:improvements}

As shown above, the median SED tends to underpredict the PAH emission.
\citet{li2001infrared} adopted a lognormal excess in the grain size distribution for the PAH component, and successfully fitted the MW PAH SED. Such an enhancement of PAH abundance does not appear in our model, suggesting that there may be additional PAH formation paths which we have not accounted for.
\citetalias{hirashita2020spectral} discussed some possible additional PAH formation paths such as stellar PAH formation \citep[e.g.][]{Latter91,Joblin:2008a,Galliano:2008a}, fragmentation by rotational disruption \citep{Hoang:2019a} and formation in dense gas \citep{Sandstrom:2010a,Chastenet:2019a}.
We emphasize again that these additional processes should not break the strong correlation
between PAH emission and metallicity
\citep[e.g.][]{Engelbracht:2005a,Draine:2007b,Hunt:2010a,Ciesla:2014a}, which is naturally explained by the strong link between dust (metal) enrichment and PAH production in our model \citep[see also][]{Seok14,Rau19}.

\citetalias{hirashita2020spectral} also argued for the necessity of a spatially resolved treatment for ISM phases, because the aromatic fraction and the grain size distribution are both sensitive to the evolution of the local physical condition of the ISM.
Thus, it would be interesting to implement our PAH evolution model directly in hydrodynamical simulations.

In this paper, we adopted dust species based on \citet{draine2007infrared}. There are other
possible sets of dust species that could explain both the MW extinction curve and IR SED
\citep[e.g.][]{Zubko:2004a,jones2013evolution,Galliano:2021a}.
\citetalias{hirashita2020spectral} also examined the dust properties (aromatic hydrocarbon materials) taken from \citet{jones2013evolution}, but as far as the comparison with the MW SED is concerned, the overall goodness of fit does not improve (or worsen) significantly.

\section{Conclusions}\label{sec:conclusions}

In this paper, we calculate the SEDs of MW analogues by post-processing the IllustisTNG simulation with our dust evolution model, which predicts the evolution of grain size distribution in a manner consistent with the chemical enrichment and merging histories of the simulated galaxies.
We selected 206 MW-like galaxies from TNG300-1 based on their stellar mass and sSFR at $z=0$.
We previously showed that this model predicts extinction curves similar to that of the MW  at $z<1$ (\citetalias{huang2021evolution}). In this study, we examine whether MW-like dust emission SEDs are also reproduced by our model of dust evolution in the cosmological assembly of MW-like galaxies.

We calculate the dust emission SED of each galaxy based on the distribution function of dust temperature under a standard MW radiation field ($U=1$).
We compare the SEDs with the MW data taken from \citet{compiegne2011global}, and find that most of the observed MW SED data points are within the 25th to 75th percentile range of the predictions, although the median underpredicts the emission at $\lambda\lesssim 10~\micron$. Our results indicate that our predicted grain size distributions are overall successful in reproducing the MW SED, while an enhancement of PAH production is still necessary for a better match.

We also examine how several quantities characterizing the galaxies affect their SEDs. We find that metallicity and dense gas fraction impact the SED. The metallicity effect is straightforward: higher metallicity leads to higher dust abundance, which raises the emission intensities at all wavelengths. The galaxy-to-galaxy dispersion in the SED is larger in the lower metallicity sample because of the greater variety in dust growth efficiency. The dense gas fraction influences the SED because it has a negative correlation with metallicity. In addition, small grains and PAHs are less abundant in dense gas because of coagulation (loss of small grains) and aliphatization (converting PAHs to non-aromatic carbon; see \citetalias{hirashita2020spectral}).
This further suppresses the emission at $\lambda\lesssim 20~\micron$ in galaxies with higher dense gas fraction.

We present our results in broader contexts at both high redshift and $z=0$. We examine the redshift evolution of MW progenitors by tracing back the main branch of the merger tree of each simulated MW analogue.
We find that the dust emission intensity increases from $z\sim 3$ to $z\sim 1$, which predominantly reflects the dust enrichment.
The present level of dust emission is established in the MW progenitors at $z\sim 1$, at which time the extinction curves also converge to a MW-like shape (\citetalias{huang2021evolution}). The SEDs have a large galaxy-to-galaxy scatter at $z\sim 3$, especially in MIR (small-grain) and PAH emission. The large variation of small-grain abundance is caused by dust growth by accretion, whose sensitivity to the metallicity, dust-to-gas ratio, and denes gas fraction produces a large variation in the abundance ratio between small and large grains.

We also compare the SED shapes at $z=0$ with those observed in nearby MW analogues, after normalizing the SEDs to the intensity at $\lambda =70~\micron$. The simulated SED shapes with $U=1$--10 broadly reproduce the observed range, but tend to underpredict the relative intensities at PAH-dominated wavelengths. Note that this underpredicting trend is also seen in the above comparison to the MW SED.

We finally test if varying parameters in our dust model can improve the above discrepancy in the predicted and observed PAH intensities at $z=0$. Assuming ionized instead of neutral PAHs does not monotonically increase the PAH emission at all wavelengths: emission from ionized PAHs is
stronger at 8~$\micron$ but weaker at 3.3 and 12~$\micron$. We also increased the ISRF intensity to $U=2$. Although higher $U$ increases the PAH emission, it overpredicts emissions at longer wavelengths.
These results imply that the underprediction of PAH feature intensity is due to the limitations of our model. We propose that including additional PAH formation paths and/or a more sophisticated spatially resolved treatment of PAH formation and evolution would be worthwhile improvements to our approach.

Our broad success in predicting the MW dust SED in this paper, together with the prediction of the MW extinction curve in \citetalias{huang2021evolution}, indicates that our model of the evolving grain size distributions in a manner consistent with the cosmological assembly of galaxies is supported by the actually observed dust properties at $z=0$. In the future, we will extend our model to other types of galaxies in the local Universe and to high-redshift galaxies.

\section*{Acknowledgements}
 
We are grateful to the anonymous referee, R. Makiya, Y.-H. Hsu, Y.-T. Lin, and D. Nelson for useful comments and discussions.
We thank the IllustrisTNG project members
for providing free access to the data used in this work.
HH thanks the Ministry of Science and Technology (MOST) for support through grant
MOST 108-2112-M-001-007-MY3, and the Academia Sinica
for Investigator Award AS-IA-109-M02. APC and CYC are supported by APC's Yushan Fellowship, awarded by the Taiwanese Ministry of Education. APC acknowledges support from the grant MOST 109-2112-M-007-011-MY3.

\section*{Data Availability}

Data directly related to this publication and its figures are available on request from
the corresponding author.
The IllustrisTNG simulations are publicly available and accessible at \url{www.tng-project.org/data} \citep{Nelson19}.




\bibliographystyle{mnras}
\bibliography{reference} 




\appendix

\bsp	
\label{lastpage}
\end{document}